\documentclass[aps,prc,preprint,showkeys,superscriptaddress,amsmath,amssymb,floatfix,nofootinbib]{revtex4-1}
\usepackage[pdftex]{graphicx}
\usepackage{dcolumn}% Align table columns on decimal point
\usepackage{bm}% bold math
\usepackage{multirow}

\def\pmb#1{\setbox0=\hbox{#1}%
  \kern-.025em\copy0\kern-\wd0 
  \kern.05em\copy0\kern-\wd0
  \kern-.025em\raise.0433em\box0 }

\def\lambdabar{\protect\@lambdabar}
\def\@lambdabar{%
\relax
\bgroup
\def\@tempa{\hbox{\raise.73\ht0
\hbox to0pt{\kern.25\wd0\vrule width.5\wd0
height.1pt depth.1pt\hss}\box0}}%
\mathchoice{\setbox0\hbox{$\displaystyle\lambda$}\@tempa}%
{\setbox0\hbox{$\textstyle\lambda$}\@tempa}%
{\setbox0\hbox{$\scriptstyle\lambda$}\@tempa}%
{\setbox0\hbox{$\scriptscriptstyle\lambda$}\@tempa}%
\egroup
}
\begin{document}

\preprint{J-PARC-TH-0313}

\title{\boldmath
Continuum $\Lambda$ spectra for a ${^6_\Lambda{\rm Li}}$ hypernucleus 
in the $^6$Li($K^-$,~$\pi^-$) reaction
%Production spectra for a ${^6_\Lambda{\rm Li}}$ hypernucleus in the ($K^-$,~$\pi^-$) and ($\pi^+$,~$K^+$) reactions on $^6$Li targets
}% Force line breaks with \\

\author{Toru~Harada}% 
\email{harada@osakac.ac.jp}
\affiliation{%
Center for Physics and Mathematics,
Osaka Electro-Communication University, Neyagawa, Osaka, 572-8530, Japan
}%Lines break automatically or can be forced with \\
\affiliation{%
J-PARC Branch, KEK Theory Center, Institute of Particle and Nuclear Studies,
High Energy Accelerator Research Organization (KEK),
203-1, Shirakata, Tokai, Ibaraki, 319-1106, Japan
}%Lines break automatically or can be forced with \\

\author{Yoshiharu~Hirabayashi}%
%\email{hirabay@iic.hokudai.ac.jp}
\affiliation{%
Information Initiative Center, 
Hokkaido University, Sapporo, 060-0811, Japan
}%Lines break automatically or can be forced with \\

\date{\today}% It is always \today, today,
             %  but any date may be explicitly specified
\begin{table}

\end{table}
\begin{abstract}
We theoretically investigate $\Lambda$ production via a ($K^-$,~$\pi^-$) reaction on a $^6$Li target, 
using the distorted-wave impulse approximation (DWIA) with a Fermi-averaged 
$K^-n \to \pi^- \Lambda$ amplitude.
We calculate $\Lambda$ production spectra using the Green's function method for 
a $^5$Li nuclear core + $\Lambda$ system, employing a $\Lambda$ folding-model potential 
based on the $^5$Li nuclear density, which is constructed from $\alpha\mbox{-}p$ 
and ${^3{\rm He}}\mbox{-}d$ cluster wave functions.
The results show that the calculated spectrum, which includes $\Lambda$ bound, resonance, 
and continuum states, agrees well with the experimental data from 
the ($K^-$,~$\pi^-$) reaction at $p_{K^-}=$ 790 MeV/$c$ (0$^\circ$), 
where substitutional $(0p,~0p^{-1})_{\Lambda n}$ 
and $(0s,~0s^{-1})_{\Lambda n}$ configurations dominate in the near-recoilless reactions.
A narrow peak corresponds to a high-lying excited state with spin-parity $J^P=$ $1^+$ at 
$E_\Lambda=$ 13.8 MeV near the ${^3{\rm He}}+d+\Lambda$ threshold, 
arising from interference effects between ${^5{\rm Li}(3/2^+)}\otimes(0s_{1/2})_\Lambda$ 
and ${^5{\rm Li}(1/2^+)}\otimes(0s_{1/2})_\Lambda$ components.
This study offers a valuable framework for extracting essential information 
on the structure and production mechanisms of hypernuclear states from experimental data.
\end{abstract}
                             % PACS, the Physics and Astronomy
                             % Classification Scheme.
%Use showkeys class option if keyword
                              %display desired
\maketitle

%-------------------------------------------------
% Text 

\section{Introduction}
\label{Intro}

Recently, the J-PARC E10 Collaboration \cite{Honda17} demonstrated 
the absence of a bound state for the neutron-rich $^6_\Lambda$H hypernucleus 
in a ($\pi^-$,~$K^+$) reaction on a $^6$Li target \cite{Harada17}. 
Additionally, the analysis of the continuum spectrum data \cite{Harada18} 
indicated that the $\Sigma^-$-$^5$He potential is repulsive, 
with a strength of $U_\Sigma=$ +30 MeV.
The J-PARC E75 Collaboration \cite{Fujioka21} has conducted a search for 
the $\Xi$ hypernucleus $^7_\Xi$H via a ($K^-$,$K^+$) reaction 
on a $^7$Li target to investigate the properties of 
the yet unsettled $\Xi$ nuclear potential. 
A search for $^6_\Xi$H using the ($K^-$,$K^+$) reaction on a $^6$Li target has 
also been proposed at J-PARC.

One of the primary objectives of this study is to elucidate the structure 
and production mechanisms of hypernuclei 
by theoretically analyzing experimental spectra 
from ($K^-$,~$\pi^-$), ($\pi^+$,~$K^+$), and ($K^-$,~$K^+$) reactions. 
As an initial step toward achieving this goal, 
we aim to establish an effective method for analyzing $\Lambda$ production spectra, 
including both bound and continuum states, 
for $\Lambda$ hypernuclei whose overall characteristics are already well understood.

\begin{figure}[tb]
\begin{center}
  \includegraphics[width=1.0\linewidth]{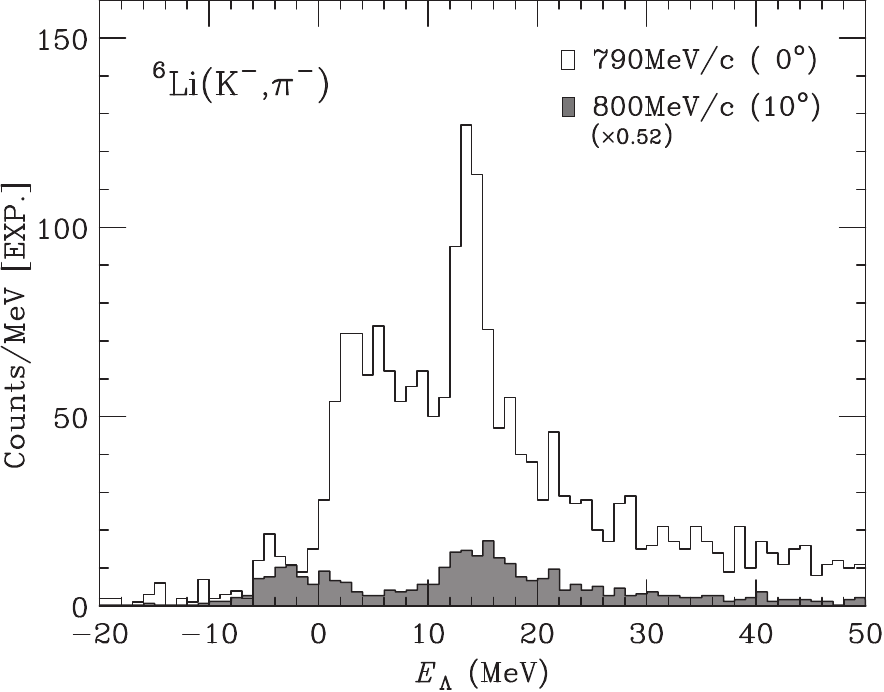}
\end{center}
\caption{\label{fig:1}
Experimental data from the $^6$Li($K^-$,~$\pi^-$) reaction at 
$p_{K^-}=$ 790 MeV/$c$ and $\theta_{\rm lab}=$ 0$^\circ$ \cite{Bertini81a},
as a function of $E_\Lambda$ measured relative to 
the ${^5{\rm Li}({\rm g.s.})}+\Lambda$ threshold, 
together with the data from the proton-coincidence nonmesonic weak decay of
${^5_\Lambda{\rm He}}$, obtained from the $\pi^-$ spectrum in a $^5_\Lambda$He lifetime 
measurement using the $^6$Li($K^-$,$\pi^-$) reaction at $p_{K^-} = 800$ MeV/$c$ \cite{Szymanski86}.
The latter dataset is normalized by multiplying a factor of 0.52 
at the $-2.5$ MeV data point for comparison.
}
\end{figure}

The ($K^-$,~$\pi^-$) reaction has played a crucial role in elucidating 
the structure of $\Lambda$ hypernuclear excited states by inducing a 
small momentum transfer of $q \simeq$ 60--80 MeV/$c$ to a $\Lambda$ 
hyperon~\cite{Feshbach66,Kerman71,Povh76}. 
This reaction has allowed the observation of prominent peaks corresponding 
to ``substitutional states'' in $\Lambda$ hypernuclei in experimental data 
on light nuclear targets \cite{Povh76,Bruckner76,Bruckner78,Bertini81a,Bertini81b}.
The Heidelberg-Saclay-Strasbourg Collaboration \cite{Bertini81a,Bertini81b} 
has reported that the $^6$Li($K^-$,$\pi^-$) reaction 
at $p_{K^-}=$ 790 MeV/$c$ and $\theta_{\rm lab}=$ 0$^\circ$ 
reveals two prominent peaks at $E_\Lambda=$ 3.8 MeV and 13.8 MeV, measured relative 
to the $^5{\rm Li}({\rm g.s})+\Lambda$ threshold, as shown in Fig.~\ref{fig:1}. 
These peaks are identified as states with spin-parity $J^p=$ $1^+$ 
in a $^6_\Lambda$Li hypernucleus, 
corresponding to substitutional 
$(0p,~0p^{-1})_{\Lambda n}$ and $(0s,~0s^{-1})_{\Lambda n}$ configurations \cite{Bertini81b}. 
This is attributed to the dominance of an angular-momentum transfer of $\Delta L =$ 0 
in the near-recoilless ($K^-$,~$\pi^-$) reactions on a $^6$Li($1^+$; g.s.) target.
In Fig.~\ref{fig:1}, we also include data from the proton-coincidence nonmesonic weak 
decay of ${^5_\Lambda{\rm He}}$, 
obtained from the $\pi^-$ spectrum in a $^5_\Lambda$He lifetime measurement 
using the $^6$Li($K^-$,$\pi^-$) reaction at $p_{K^-} = 800$ MeV/$c$ \cite{Szymanski86}.
Notably, the small yield from $^6_\Lambda{\rm Li}(1^+)$ at 13.8 MeV is 
comparable to that from $^6_\Lambda{\rm Li}({\rm g.s.})$ at $-$4.5 MeV. 
These data suggest a minimal ($\sim$4\%) admixture of 
($\alpha\mbox{-}p$)-$\Lambda$ to ($^3{\rm He}\mbox{-}d$)-$\Lambda$ in $^6_\Lambda$Li, 
whose wave functions exhibit the spatial symmetries of $[f]=$ [411] and [321] 
in the Young's scheme \cite{Neudatchin91}, respectively, 
as discussed by Majling {\it et al}.~\cite{Majling80}.

%%%%%%%%%%%%%%%%%%%%%%%%%%%%%%%%%%%%%%%%%%%%%%%%%%%%%%%
%\Figuretable{FIG. 1}
% Figure 1
%%%%%%%%%%%%%%%%%%%%%%%%%%%%%%%%%%%%%%%%%%%%%%%%%%%%%%%

Many authors \cite{Auerbach83,Bando90,Gal16}
have theoretically studied spectroscopy of $p$-shell $\Lambda$ hypernuclei
in ($K^-$,~$\pi^-$), ($\pi^+$,~$K^+$), and ($e$,~$e'K^+$) reactions, 
using a distorted-wave impulse approximation.
Shell-model \cite{Majling80,Millener07} and 
cluster-model \cite{Motoba83,Motoba85,Ohkura93} descriptions
discussed the structure and production of $^6_\Lambda$Li 
via the $^6$Li($K^-$,~$\pi^-$) reaction within a bound-state approximation. 
Auerbach and Van Giai \cite{Auerbach80} studied basic gross features of 
continuum effects considering $\Lambda N^{-1}$ configurations in 
the ($K^-$,~$\pi^-$) reaction on $^6$Li, together with $^7$Li and $^9$Be. 

In this paper, 
we theoretically investigate $\Lambda$ production 
via the ($K^-$,~$\pi^-$) reaction on a $^6$Li target, 
using the distorted-wave impulse approximation 
with the Fermi-averaged $K^-n \to \pi^- \Lambda$ amplitudes 
incorporating nuclear medium effects.
To describe $\Lambda$ bound, resonance, and continuum states for $^6_\Lambda$Li, 
we calculate a $\Lambda$ production spectrum by employing 
the Green's function method \cite{Morimatsu94}, 
which provides a comprehensive description of $\Lambda$-nucleus dynamics 
to potential interactions.
We obtain the $\Lambda$ folding-model potentials with 
the $^5$Li nuclear densities based on $\alpha\mbox{-}p$ and  
${^3{\rm He}}\mbox{-}d$ cluster wave functions. 
We examine the shape and magnitude of the $\Lambda$ production spectra, 
comparing them with the experimental data \cite{Bertini81a}. 
This study aims to extract essential information on the structure 
and production mechanisms of $^6_\Lambda$Li from the data of 
the $^6$Li($K^-$,~$\pi^-$) reaction.

\section{Method}
\label{Calc}

\subsection{Distorted wave impulse approximation}

The double-differential cross section for $\Lambda$ production 
on a nuclear target in the ($K^-$,~$\pi^-$) reaction is often calculated using 
the distorted wave impulse approximation (DWIA) \cite{Hufner74,Auerbach83}.
According to the Green's function method \cite{Morimatsu94}, 
the double-differential laboratory cross section  on a nuclear target 
with spin $J_A$ and its $z$-component $M_A$ \cite{Hufner74}
can be expressed as follows (in natural units, $\hbar = c = 1$):
\begin{equation}
{{d^2\sigma} \over {d\Omega dE}} 
 = {1 \over {[J_A]}} \sum_{M_A}S(E_B), 
\label{eqn:e1}
\end{equation}
where $[J_A] = 2J_A+1$, and $S(E_B)$ is the strength function, which is given by
\begin{eqnarray}
S(E_B)&=&-{1 \over \pi}{\rm Im} \sum_{c c'}
\int d{\bm r}d{\bm r'}
F^{\Lambda \dagger}_{c}({\bm r}) 
{G}_{cc'}(E_B;{\bm r},{\bm r}')\nonumber\\
&& \times 
F^{\Lambda}_{c'}({\bm r}'),
\label{eqn:e2}
\end{eqnarray}
as a function of the energy $E_B$ for hypernuclear final states.
Here, ${G}_{cc'}$ is a coupled-channel (CC) Green's function \cite{Harada00} 
that accounts for spin couplings between hypernuclear states, 
where $c$ ($c'$) denotes a complete set of eigenstates of the system. 
The $\Lambda$ production amplitude $F^{\Lambda}_{c}$ is defined as
\begin{equation}
  F^{\Lambda}_{c} = 
  \beta^{1 \over 2}{\overline{f}}_{K^-n \to \pi^-\Lambda}
  \chi_{{\bm p}_{\pi}}^{(-) \ast}
  \chi_{{\bm p}_{K}}^{(+)} 
  \langle c \, | \hat{\mit\psi}_n | \, \Psi_A \rangle,
\label{eqn:e3}
\end{equation}
where 
$\chi_{{\bm p}_{\pi}}^{(-)}$ and $\chi_{{\bm p}_{K}}^{(+)}$ 
are the distorted waves for the outgoing $\pi^-$ and incoming $K^-$ mesons, respectively.
The function $\langle c \, | \hat{\mit\psi}_n  | \, \Psi_A \rangle$ represents 
the hole-state wave function for a struck neutron in the target.
The quantity ${\overline{f}}_{K^-n \to \pi^-\Lambda}$ is the Fermi-averaged amplitude
for the $K^-n \to \pi^-\Lambda$ reaction in the nuclear medium \cite{Harada04,Harada22}, 
and $\beta$ is a kinematical factor converting from the two-body $K^-$-nucleon ($N$) 
laboratory system to the $K^-$-nucleus laboratory system \cite{Dover83,Koike08}.
The energy and momentum transfer are given by
\begin{eqnarray}
\omega   =  E_{K}-E_{\pi}, 
\qquad {\bm q}  =  {\bm p}_{K} - {\bm p}_{\pi}, 
\label{eqn:e4}
\end{eqnarray}
where $E_{K} = ({\bm p}_{K}^2+m^2_{K})^{1/2}$ and 
$E_{\pi} = ({\bm p}_{\pi}^2+m^2_{\pi})^{1/2}$.
Here, ${\bm p}_K$ and $m_K$ (${\bm p}_\pi$ and $m_\pi$) 
denote the laboratory momenta and rest masses of $K^-$ and $\pi^-$, respectively.

The distorted waves, $\chi_{{\bm p}_{\pi}}^{(-)}$ and $\chi_{{\bm p}_{K}}^{(+)}$,
in Eq.~(\ref{eqn:e3}) are estimated using the eikonal approximation, 
with total cross sections of $\sigma_K = 30$ mb for $K^- N$ and 
$\sigma_\pi = 32$ mb for $\pi^- N$, and distortion parameters 
$\alpha_K = \alpha_\pi = 0$ \cite{Harada04}. 
Recoil effects are also incorporated, leading to an effective momentum transfer 
$q_{\rm eff} \simeq (1-1/A)q \simeq 0.83 q$ for a light nuclear 
system with $A = 6$.

\subsection{Model wave functions}

%%%%%%%%%%%%%%%%%%%%%%%%%%%%%%%%%%%%%%%%%%%%%%%%%%%%%%%
%Table 1
\begin{table}[t]
\caption{\label{tab:table1}
Calculated and input values of neutron $0p^{-1}$ and $0s^{-1}$ states 
for $^6{\rm Li}(1^+;{\rm g.s.})$. 
$\epsilon_N$ and $\langle r^2 \rangle^{1/2}$ denote energies and 
rms radii for the neutron-hole states, respectively, 
which are given by spectroscopic amplitudes in a $\alpha+d$ cluster model. 
The width $\Gamma$ and spectroscopic factor $S_{\ell_N}$ for these states 
are taken from the experimental data \cite{Tilley02,Jacob66}. 
}
\begin{ruledtabular}
\begin{tabular}{ccccc}
\noalign{\smallskip}
State  &  $\epsilon_N$ &$\langle r^2 \rangle^{1/2}$  & $\Gamma$  & $S_{\ell_N}$   \\
\noalign{\smallskip} 
       &    (MeV)     & (fm)   &(MeV)                           &                \\
\noalign{\smallskip}\hline\noalign{\smallskip}
$0p^{-1}$  & $-$4.5    &  3.58  &  1.2\footnotemark[1]    &  0.8\footnotemark[2]  \\
$0s^{-1}$  & $-$21.3   &  2.34  &  0.2\footnotemark[1]    &  1.6\footnotemark[2]  \\ 
\end{tabular}
\end{ruledtabular}
\footnotetext[1]{
Ref.~\cite{Tilley02}.}
\footnotetext[2]{
Ref.~\cite{Jacob66}.}
\end{table}
%%%%%%%%%%%%%%%%%%%%%%%%%%%%%%%%%%%%%%%%%%%%%%%%%%%%%%%

For a  $^{6}$Li($1^+;{\rm g.s.}$) target, 
the wave function is expressed in the $\alpha +d$ cluster model \cite{Fujiwara80},
\begin{eqnarray}
\Psi_{A}^{(^6{\rm Li})}
&=& {\cal A}\bigl[[\phi_{\alpha}\phi_{d}]_{1^+}
\otimes\chi_{L_A}^{(\alpha\mbox{-}d)}\bigr]_{J_A}, 
\label{eqn:e5}
\end{eqnarray}
where ${\cal A}$ is an antisymmetrization operator, 
$\phi_{\alpha}$ and $\phi_{d}$ are the internal wave functions of $\alpha$ and $d$, 
respectively, and $\chi_{L_A}^{(\alpha\mbox{-}d)}$ is the relative wave function
between $\alpha$ and $d$. 
Here, we use a $(0s)^4$ harmonic oscillator wave function for $\phi_{\alpha}$ 
with a size parameter $b_\alpha=$ 1.39 fm, 
and a $(0s)^2$ wave function for $\phi_{d}=c_1\phi_1+c_2\phi_2$, 
where $\phi_i=\exp{\{-(r/2b_i)^2\}}$ with $c_1=$ 0.07197, $c_2=$ 0.05090, 
$b_1=$ 1.105 fm, and $b_2=$ 2.395 fm, incorporating the distortion effect 
in $d$ \cite{Kanada82}. 
According to the orthogonality condition model (OCM) \cite{Fujiwara80}
with an appropriate $\alpha\mbox{-}d$ potential \cite{Sakuragi86},
we obtain the relative wave function $\chi_{L_A}^{(\alpha\mbox{-}d)}$ 
with an orbital angular momentum of $L_A=0$ 
and a binding energy of 1.47 MeV relative to the $\alpha+d$ threshold. 
The charge root-mean-square (rms) radius of $^6$Li(1$^+;{\rm g.s.}$) is obtained as 2.53 fm, 
which is in good agreement with the experimental value of $2.56 \pm 0.05$ fm 
from electron elastic scatterings \cite{Vries87}. 

The wave function 
$\langle c \, | \hat{\mit\psi}_n  | \, \Psi_A \rangle$ 
for a neutron $p$-hole or $s$-hole state in Eq.~(\ref{eqn:e3}) 
can be determined as a spectroscopic amplitude describing the $\ell_N$-neutron 
pickup from $^{6}$Li$(1^+;{\rm g.s.})$ in the $\alpha + d$ cluster model.
Table~\ref{tab:table1} lists the values of the input energies $\epsilon_N$ and 
widths $\Gamma$, as well as the spectroscopic factors $S_{\ell_N}$ 
for the neutron $0p^{-1}$ and $0s^{-1}$ orbits, 
based on the experimental data from the $^{6}$Li($p$,~2$p$) 
reaction \cite{Tilley02,Jacob66}. 

For $^6_\Lambda$Li hypernuclear final states, 
we employ multi-configurational wave functions in a
$^5$Li nuclear core + $\Lambda$ model, 
which are expressed as
\begin{eqnarray}
\Psi^{(^6_\Lambda{\rm Li})}_{J_B}
&=& \sum_{c}
\bigl[\Phi^{(^5{\rm Li})}_{J_C}\otimes
\varphi^{(\Lambda)}_{j \ell s}({\bm r})\bigr]_{J_B}, 
\label{eqn:e6}
\end{eqnarray}
where the abbreviation $c=\{J_C j \ell s \}$, 
$\Phi^{(^5{\rm Li})}_{J_C}$ denotes the wave function for 
$^5$Li with spin $J_C$,  $\varphi^{(\Lambda)}_{j \ell s}$ represents 
the relative wave function with the $(j \ell s)$ state, 
and ${\bm r}$ is the relative coordinate between $^5$Li($J_C$) and $\Lambda$.
Since non-spin-flip processes predominantly occur in $K^- n \to \pi^- \Lambda$ 
reactions at $p_{K^-}=$ 700--800 MeV/$c$,
we consider hypernuclear final states of $^6_\Lambda$Li 
with $J^P=$ $(1^+ \otimes \Delta L)=$ 
1$^+$, 0$^-$, 1$^-$, 2$^-$, 1$^+$, 2$^+$, 3$^+$, $\cdots$, 
which correspond to configurations of ${^5{\rm Li}(J_C)} \otimes \Lambda$,
where $\Delta L=$ 0, 1, 2, $\cdots$, as the angular momentum transfer to 
the $^6{\rm Li}$(1$^+;{\rm g.s.}$) target in the ($K^-$,~$\pi^-$) reaction.

\begin{figure}[t]
  % fig1
  \begin{center}
  \includegraphics[width=1.0\linewidth]{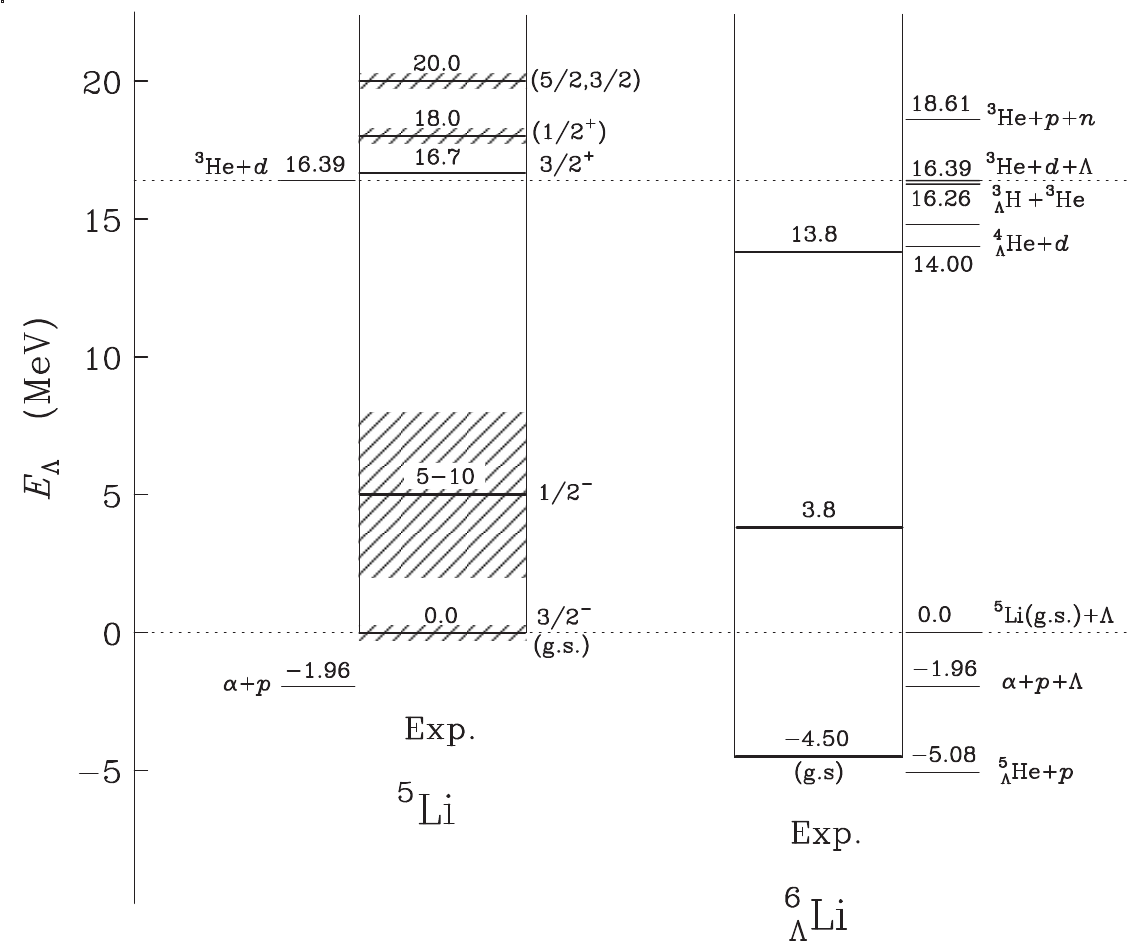}
  \end{center}
  \caption{\label{fig:2}
Experimental energy levels of $^6_\Lambda$Li and the related decay thresholds, 
as a function of $E_\Lambda$ measured from the $^5{\rm Li}({\rm g.s.)}+\Lambda$ threshold, 
together with those of $^5$Li as a core nucleus \cite{Ajzenberg88}. 
The data are taken from Ref.~\cite{Bertini81a}. 
}
\end{figure}

Figure~\ref{fig:2} presents the experimental energy levels and associated decay 
thresholds of $^6_\Lambda$Li \cite{Bertini81a}, 
together with those of $^5$Li \cite{Ajzenberg88}.
In this study, we measure the $\Lambda$ energy $E_\Lambda$ measured relative to 
the $^5{\rm Li}({\rm g.s.})+\Lambda$ threshold, which is defined as
\begin{eqnarray}
E_\Lambda  = E_B-M({^5{\rm Li}}({\rm g.s.}))-m_\Lambda = -B_\Lambda,
\label{eqn:e7}
\end{eqnarray}
where $M({^5{\rm Li}}({\rm g.s.}))$ and $m_\Lambda$ are 
the masses of $^5{\rm Li}({\rm g.s.})$ and $\Lambda$, respectively.
We note that a low-lying narrow $3/2^-$ resonance state of 
$^5{\rm Li}({\rm g.s.})$ exists at ($E_r$,~$\Gamma$) = (1.96 MeV, 0.65 MeV) 
above the $\alpha + p$ threshold, while a broad state 
of $^5{\rm Li}(1/2^-)$ is located at 5--10 MeV \cite{Tilley02}, 
corresponding to the valence $0p_{3/2}$ and $0p_{1/2}$ proton orbits  
in the $\alpha\mbox{-}p$ configurations, respectively.
On the other hand, high-lying resonance states, 
$^5{\rm Li}(3/2^+)$ at 16.7 MeV
and 
$^5{\rm Li}(1/2^+)$ at 18.0 MeV,
near the $^3{\rm He}+d$ threshold,
are expected to be described by the $^3{\rm He}\mbox{-}d$ configurations.

%\Figuretable{FIG. 2}
% Figure 2
%%%%%%%%%%%%%%%%%%%%%%%%%%%%%%%%%%%%%%%%%%%%%%%%%%%%%%%%%%%%%%%%%
%%%%%%%%%%%%%%%%%%%%%%%%%%%%%%%%%%%%%%%%%%%%%%%%%%%%%%%

\subsection{Coupled-channel calculations}

To investigate $\Lambda$-nucleus dynamics within this model space, 
we obtain the wave functions $\varphi^{(\Lambda)}_{j \ell s}$ in Eq.~(\ref{eqn:e6}) 
by solving the following coupled-channel (CC) equation 
with a $\Lambda$-nucleus optical potential $\hat{U}_\Lambda$: 
\begin{eqnarray}
& & \Big[ -{1 \over 2\mu_c}\nabla^2
+U^\Lambda_{cc}({\bm r}) - (E_\Lambda - {\varepsilon}_c) \Big]
\varphi^{(\Lambda)}_{j \ell s}({\bm r}) \nonumber\\
& & =- \sum_{c \neq c'}U^\Lambda_{cc'}({\bm r})
\varphi^{(\Lambda)}_{j' \ell' s'}({\bm r}),
\label{eqn:e7}
\end{eqnarray}
where ${\varepsilon}_c$ and $\mu_c$ denote a channel energy and a 
reduced mass of the ${^5{\rm Li}(J_C)}$-$\Lambda$ system for the $c$ channel, 
respectively. 
The matrix elements $U^{\Lambda}_{cc'}$ can be systematically evaluated using 
Racah algebra \cite{Davies64,Tamura65,Glendenning83}:
\begin{eqnarray}
U^{\Lambda}_{cc'}(r) 
&=&
\left\langle 
[\Phi^{(^5{\rm Li})}_{J_C}, {\cal Y}^{(\Lambda)}_{j \ell s}\bigr]_{J_B}
\middle|\hat{U}_\Lambda 
\middle|
[\Phi^{(^5{\rm Li})}_{J'_C}, {\cal Y}^{(\Lambda)}_{j' \ell' s'}\bigr]_{J_B}
\right\rangle \nonumber\\
&=& 
\sum_{LSK}\,C^{J_B}_{LSK}(J_CJ'_C){\cal F}^{J_CJ'_C}_{LSK}(r),
\label{eqn:e12}
\end{eqnarray}
where ${\cal Y}^{(\Lambda)}_{j \ell s}=[{Y}_\ell \otimes X_s]_j$ 
is the spin-orbit function for $\Lambda$, 
$C^{J_B}_{LSK}(J_CJ'_C)$ is a purely geometrical strength factor \cite{Glendenning83}, 
and ${\cal F}^{J_CJ'_C}_{LSK}(r)$ is the nuclear form factor for 
the $(\alpha\mbox{-}p)\mbox{-}\Lambda$ or $({^3{\rm He}}\mbox{-}d)\mbox{-}\Lambda$ 
system \cite{Davies64,Tamura65}.
To compute the CC Green's function $G_{cc'}$ in Eq.~(\ref{eqn:e2}), 
we numerically solve the associated CC equation for $G_{cc'}$, which is given by
\begin{eqnarray}
G_{cc'}(E_B) 
&=& G^{(0)}_{c}(E_B)\delta_{cc'}                 \nonumber \\
&+& \sum_{c''}G^{(0)}_{c}(E_B)U^\Lambda_{c c''}G_{c''c'}(E_B),
\label{eqn:e10}
\end{eqnarray}
where $G^{(0)}_{c}$ is the free Green's function for the $c$ channel.

\subsection{\boldmath
Fermi-averaged $\bm{K^-n\to\pi^-\Lambda}$ amplitudes}

%%%%%%%%%%%%%%%%%%%%%%%%%%%%%%%%%%%%%%%%%%%%%%%%%%%%%%%%%%%%%%%%%%%%%%%%%%%%%%%%%
%\Figuretable{FIG. 2}
% Figure 2
\begin{figure}[t]
\begin{center}
  \includegraphics[width=1.0\linewidth]{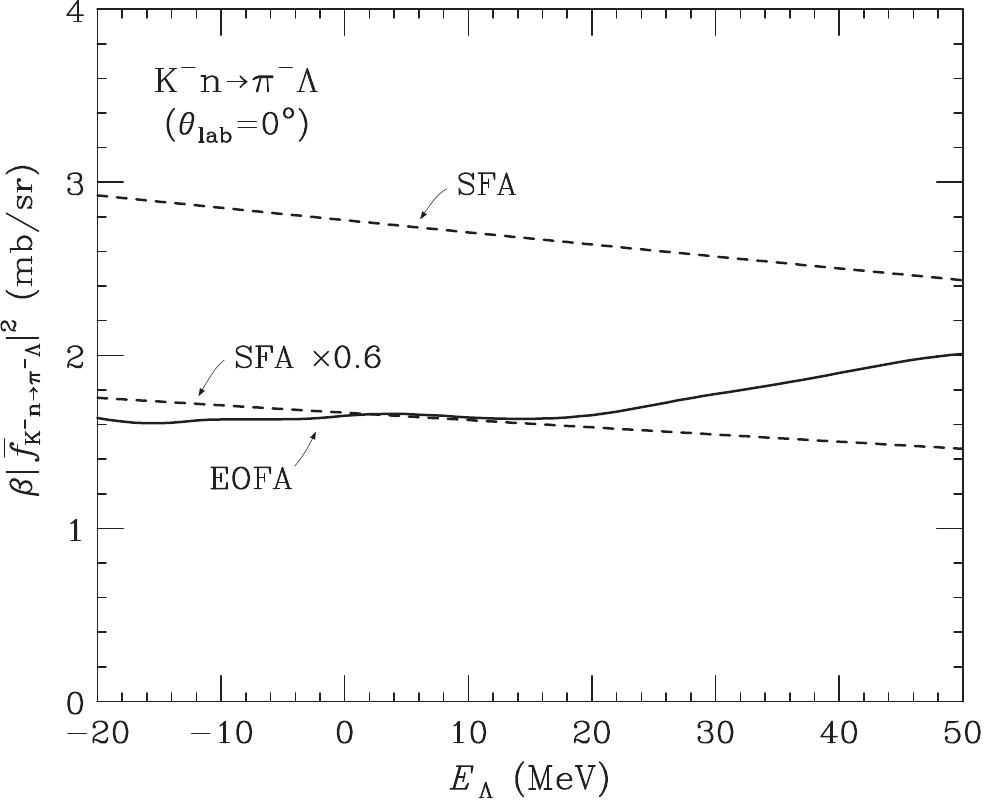}
\end{center}
\caption{\label{fig:3}
Fermi-averaged $K^-n\to\pi^-\Lambda$ differential cross sections 
of $\beta|\overline{f}_{K-n\to\pi^-\Lambda}|^2$
at $p_{K^-}=$ 790 MeV/$c$ and $\theta_{\rm lab}=$ 0$^\circ$,
as a function of $E_\Lambda$ measured relative to the 
${^5{\rm Li}({\rm g.s.})}+\Lambda$ threshold.
The solid and dashed curves denote the values calculated by EOFA and SFA, respectively, 
using the elementary $K^-n\to\pi^-\Lambda$ amplitudes given 
by Gopal {\it et al}.~\cite{Gopal77}.
}
\end{figure}
%%%%%%%%%%%%%%%%%%%%%%%%%%%%%%%%%%%%%%%%%%%%%%%%%%%%%%%%%%%%%%%%%%%%%%%%%%%%%%%%%%

In the DWIA calculations, 
considering nuclear medium effects is crucial 
in nuclear ($K^-$,~$\pi^-$) reactions \cite{Zofka84,Rosenthal80} 
because narrow $Y^*$ resonances serve as intermediate states 
in the $K^-n \to \pi^-\Lambda$ processes.
The ``in-medium'' differential laboratory cross sections 
for the $K^-n \to \pi^-\Lambda$ reactions in the nucleus 
are approximately estimated as
\begin{eqnarray}
\left(\frac{d\sigma}{d\Omega}\right)_{\rm lab}
=\beta |\overline{f}_{K^-n \to \pi^-\Lambda}|^2, 
\label{eqn:e8}
\end{eqnarray}
where $\overline{f}_{K^-n \to \pi^-\Lambda}$ represents 
the Fermi-averaged $K^-n \to \pi^-\Lambda$ amplitude. 
Note that nuclear ($K^-$,~$\pi^-$) reactions at 
$p_{K^-} \simeq 700$--$800$ MeV/$c$ and $\theta_{\rm lab} \simeq 0^\circ$ 
occur under near-recoilless conditions, characterized by a small momentum 
transfer of $q \simeq 60$--$80$ MeV/$c$ \cite{Feshbach66,Kerman71}. 
We have difficulties satisfying the on-energy-shell $K^-n \to \pi^-\Lambda$ 
processes in the optimal Fermi-averaging (OFA) \cite{Harada04}
to evaluate the values of $\overline{f}_{K^-n \to \pi^-\Lambda}$ 
using the elementary $K^-n \to \pi^-\Lambda$ amplitudes provided 
by Gopal {\it et al}.~\cite{Gopal77}. 
To overcome the difficulties, therefore, 
we must require the extended optimal Fermi-averaging (EOFA) method \cite{Harada22}
in this calculation.  

Figure~\ref{fig:3} shows the Fermi-averaged $K^-n\to\pi^-\Lambda$ differential cross 
sections of $\beta |\overline{f}_{K^-n \to \pi^-\Lambda}|^2$ 
obtained using EOFA at 790 MeV/$c$ and $\theta_{\rm lab}=$ 0$^\circ$ \cite{Harada22}, 
as a function of $E_\Lambda$, along with results from 
the standard Fermi-averaging (SFA) \cite{Rosenthal80}, which incorporates 
nucleon binding effects. 
The energy dependence of $\beta |\overline{f}_{K^-n \to \pi^-\Lambda}|^2$
is relatively weak, 
while the magnitude of EOFA is almost half that of SFA.
This result confirms that EOFA is essential for resolving 
discrepancies between theoretical and experimental angular distributions 
for $^{12}{\rm C}(K^-,~\pi^-){^{12}_\Lambda{\rm C}}$ reactions 
at $p_{K^-} =$ 800 MeV/$c$ \cite{Chrien79,Dover79}.
Consequently, we expect that 
the EOFA-derived Fermi-averaged $K^-n\to\pi^-\Lambda$ amplitudes 
will also be applicable to $^6$Li($K^-$,~$\pi^-$) reactions.

\section{\boldmath
${\bm \Lambda}$-nucleus potentials}
\label{potential}

%%%%%%%%%%%%%%%%%%%%%%%%%%%%%%%%%%%%%%%%%%%%%%%%%%%%%%%%%%%%%%%%%%%%%%%%%%%%%%%%%
%\Figuretable{FIG. 4}
% Figure 4
\begin{figure}[tb]
\begin{center}
  \includegraphics[width=1.0\linewidth]{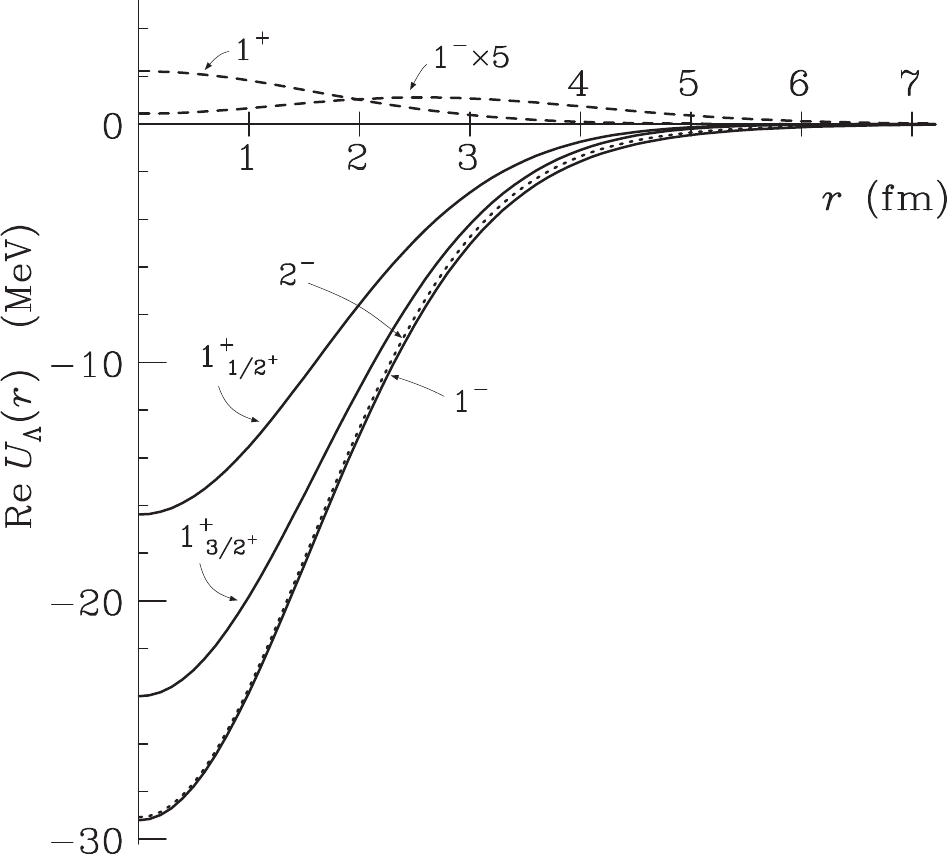}
\end{center}
\caption{\label{fig:4}
Real parts of $\Lambda$ folding-model potentials for several spin $J_B$ states 
in the ${^5{\rm Li}}\mbox{-}\Lambda$ system, 
as a function of the relative distance $r$ between the center of mass of 
${^5{\rm Li}}(J_C)$ and the $\Lambda$.
Solid curves denote the diagonal potentials for 
$J_B=$ $1^-$ in ($\alpha\mbox{-}p$)-$\Lambda$ channels 
and for $J_B=$ $1^+$ in ($^3{\rm He}\mbox{-}d$)-$\Lambda$ channels; 
a dotted curve for $J_B=$ $2^-$ in ($\alpha\mbox{-}p$)-$\Lambda$ channels. 
Dashed curves denote the coupling potentials for $J_B=$ $1^-$ and $1^+$. 
}
\end{figure}
%%%%%%%%%%%%%%%%%%%%%%%%%%%%%%%%%%%%%%%%%%%%%%%%%%%%%%%%%%%%%%%%%%%%%%%%%%%%%%%%%%

To examine $\Lambda$-nucleus dynamics, 
the CC Green's function $G_{c'c}$ is obtained by numerically 
solving the CC equation with 
a $\Lambda$-nucleus (optical) potential $\hat{U}_{\Lambda}$, 
which is expressed as
\begin{eqnarray}
\hat{U}_\Lambda({\bm r}) &=& 
U^\Lambda_0({\bm r}) + U^\Lambda_1({\bm r}){\bm I}\cdot{\bm S}, 
\label{eqn:e9}
\end{eqnarray}
where ${\bm I}$ and ${\bm S}$ denote the spin operators for the $^{5}$Li 
nuclear core and the $\Lambda$, respectively, and $U^\Lambda_s({\bm r})$ 
is obtained using a folding-model potential procedure, 
\begin{eqnarray}
U^\Lambda_s({\bm r}) &=& 
\int \bar{v}^{(s)}_{\Lambda N}({\bm r}-{\bm r}')
\rho^{(^5{\rm Li})}_{s}({\bm r}')d{\bm r}', 
\label{eqn:e10}
\end{eqnarray}
where $\bar{v}^{(s)}_{\Lambda N}$ and $\rho^{(^5{\rm Li})}_s$ denote 
the $\Lambda N$ effective interaction and the $^5$Li nuclear density
for the non-spin ($s=0$) or spin ($s=1$) state, respectively.
We omit the spin-orbit ${\bm L}\cdot{\bm S}$ term, 
as it is well known that the $\Lambda$ spin-orbit potential 
is small \cite{Bando90,Gal16}. 

The density $\rho^{(^5{\rm Li})}_s$ is constructed from the 
$\alpha$-$p$ or ${^3{\rm He}}$-$d$ cluster wave function \cite{Fujiwara80},
reproducing the charge radii of $^5{\rm Li}({\rm g.s.})$, $^3$He, and $d$.
When the nuclear shrinkage effect due to the $\Lambda$ is enhanced 
in the ${^3{\rm He}}$-$d$ system,
the distance $D$ between the ${^3{\rm He}}$ and $d$ clusters is expected to decrease
upon the addition of a $\Lambda$; the value of $D$ should be adjusted to fit 
the experimental data 
of $^6_\Lambda{\rm Li}(1^+)$ from the ($K^-$,~$\pi^-$) spectrum. 
Here, we take $D = 1.8$ fm using the ${^3{\rm He}}$-$d$ cluster 
potentials \cite{Rocca17}, which corresponds to a rms radius 
of $\langle R^2 \rangle^{1/2}=$ 2.18 fm between the ${^3{\rm He}}$ 
and $d$ clusters, 
compared with $\langle R^2 \rangle^{1/2}=$ 2.44 fm and 2.02 fm 
for $1^+_1$ and $1^+_2$ states of $^6_\Lambda{\rm Li}$, respectively, 
obtained in the ${^3{\rm He}} + d + \Lambda$ three-body model \cite{Ohkura93}.

The $\Lambda N$ effective interaction for this model space is represented by 
a single Gaussian form \cite{Motoba83,Motoba85,Ohkura93}:
\begin{eqnarray}
{v}_{\Lambda N}({\bm r})
=v^0_{\Lambda N}(1+\eta {\bm \sigma}_N\cdot{\bm \sigma}_\Lambda)
\exp{\left\{-({\bm r}/b)^2\right\}}, 
\label{eqn:e11}
\end{eqnarray}
with $v^0_{\Lambda N}=$ $-21.19$ MeV, $b=$ 1.49 fm, and $\eta=$ $-$0.072. 
The range parameter $b$ is determined to reproduce the $\Lambda N$ scattering data 
at low energies~\cite{Dalitz72}, 
while the potential strength $v^0_{\Lambda N}$ is adjusted to 
reproduce the $\Lambda$ binding energy of 4.50 MeV \cite{Bertini81a} 
for $^6_\Lambda{\rm Li}({\rm g.s.})$ within the folding-model potential.
The parameter $\eta$ implies that the spin-singlet $\Lambda N$ state ($S_{<}$) 
is more attractive than the spin-triplet $\Lambda N$ state ($S_{>}$)
\cite{Motoba83,Motoba85,Ohkura93,Hiyama96,Yamamoto10}.
Using folding-model potentials with a size parameter of $b_{N} = 1.61$~fm, 
we calculated the $\Lambda$ binding energies ($B_\Lambda$) to 
be 1.90~MeV for $^4_\Lambda{\rm He}(0^+)$, 1.25~MeV for $^4_\Lambda{\rm He}(1^+)$, 
and 0.04~MeV for $^3_\Lambda{\rm H}(1/2^+)$, 
compared with the corresponding experimental values 
of $2.39 \pm 0.07$~MeV, $0.94 \pm 0.04$~MeV, 
and $0.13 \pm 0.09$~MeV \cite{Juric73}, respectively. 
%%%%%%%%%%
The parameters of the $\Lambda N$ effective interaction in  
Eq.~(\ref{eqn:e11}) are adjusted to reproduce the experimental binding  
energy of $^6_\Lambda$Li (g.s.), thereby incorporating the main effects  
of dispersion, Pauli blocking, rearrangement, and $\Lambda$-$\Sigma$  
coupling within the two-body $YN$ dynamics.  
The energy dependence of the $\Lambda$-nucleus potential,
arising from an effective mass $\mu^*$~\cite{Harada23}, 
is neglected here for simplicity.  
Since the three-body $YNN$ force arising from 
$\Lambda N$-$\Sigma N$ coupling is not explicitly included,  
the fine structure of the binding energies of  
$^4_\Lambda{\rm He}(0^+)$, $^4_\Lambda{\rm He}(1^+)$,  
and $^3_\Lambda{\rm H}(1/2^+)$ is not fully reproduced.  
Nevertheless, the present folding-model approach remains sufficient for a  
phenomenological analysis of $\Lambda$ production, as shown in  
Sec.~\ref{Results}.
%%%%%%%%%%%%%%

Figure~\ref{fig:4} presents the real parts of the $\Lambda$ 
folding-model potentials for $^6_\Lambda$Li with $J_B=$ $1^-$ and $2^-$
in the ($\alpha\mbox{-}p$)-$\Lambda$ channels, 
as well as for $J_B=$ $1^+$, which includes 
$^5{\rm Li}(3/2^+)\otimes (0s_{1/2})_\Lambda$ and 
$^5{\rm Li}(1/2^+)\otimes (0s_{1/2})_\Lambda$ 
components in the ($^3{\rm He}\mbox{-}d$)-$\Lambda$ channels. 
The shapes of these potentials are very similar, but their strengths 
differ moderately due to the ${\bm I}\cdot{\bm S}$ term in Eq.~(\ref{eqn:e9}). 
For $1^-$, the coupling term is small 
because the valence $0p_{1/2}$ proton in $^5{\rm Li}(1/2^-)$ 
is far from the $\alpha$ core in the $(\alpha\mbox{-}p)\mbox{-}\Lambda$ system. 
For $1^+$, on the other hand, the $\Lambda$ diagonal 
potential for $^5{\rm Li}(3/2^+)$ is more attractive than that for
$^5{\rm Li}(1/2^+)$ because 
$U_1^\Lambda \langle {\bm I}\cdot{\bm S}\rangle$ is attractive 
(repulsive) for the former (latter), 
where $ U_1^\Lambda  > 0$ and  
$\langle{\bm I}\cdot{\bm S}\rangle = J_B(J_B+1)-J_C(J_C+1)-3/4$. 
This result mainly stems from the nature of the $\Lambda N$ effective interaction. 
Therefore, we confirm that the spin effects arising from 
the ${\bm I}\cdot{\bm S}$ term play a crucial role in 
the $\Lambda$ fine structure and spectroscopy of 
$^6_\Lambda{\rm Li}$ \cite{Motoba83,Motoba85,Hiyama96}.

\section{Results}
\label{Results}

\subsection{\boldmath
Energy levels of $^6_\Lambda$Li}
\label{levels}

%%%%%%%%%%%%%%%%%%%%%%%%%%%%%%%%%%%%%%%%%%%%%%%%%%%%%%%
%Table 2
\begin{table*}[hbt]
\caption{\label{tab:table2}
Energies and widths of the $\Lambda$ bound and resonance states of $^6_\Lambda$Li
in the $^5{\rm Li}(J_C)\mbox{-}\Lambda$ system, 
together with the experimental data \cite{Bertini81a}.
These values are obtained from pole positions of the $S$-matrix in the complex $E$ plane.
The energies in parentheses are measured from the ${^3{\rm He}}+d+\Lambda$ 
threshold.
}
\begin{ruledtabular}
\begin{tabular}{clcccccclc}
\noalign{\smallskip}
State & &\multicolumn{2}{c}{Configurations} && $E^{\rm cal}_\Lambda$  &  $\Gamma^{\rm cal}$  
& Sheet &
            \multicolumn{1}{c}{$E^{\rm exp}_\Lambda$}  &  $\Gamma^{\rm exp}$  \\
\noalign{\smallskip}
       \cline{3-4}  
\noalign{\smallskip}
 $J_B^P$ &&   $^5{\rm Li}(J_C)$ &   $(n \ell)_{j_\Lambda}$  &&  (MeV) &  (MeV) &     & 
                                                                (MeV) &  (MeV)    \\
\noalign{\smallskip}\hline\noalign{\smallskip}
\multicolumn{3}{l}{($\alpha$-$p$)-$\Lambda$} \\
$1^-$   &&  $3/2^-$, $1/2^-$  &  $0s_{1/2}$  &&  $-$4.51 & 1.2     & [$+$$+$] & \multirow{2}{*}{$\Big\}$\,$-$4.50} & \\
$2^-$   &&  $3/2^-$           &  $0s_{1/2}$  &&  $-$4.26 & 1.2     & [$+$] &          &        \\
$1^+_1$   &&  $3/2^-$, $1/2^-$  &  $0p_{3/2}$, $0p_{1/2}$ && $-$0.45 & 0.98 & [$-$$-$] & \multirow{2}{*}{$\Big\}$\,\ \ \ 3.8} & \\
$1^+_2$   &&  $3/2^-$, $1/2^-$  &  $1p_{3/2}$, $1p_{1/2}$ && 3.1  & 5.6  & [$-$$-$]   &          &        \\
\multicolumn{3}{l}{($^3{\rm He}$-$d$)-$\Lambda$} \\
$1^+_1$  &&  $3/2^+$, $1/2^+$  &$0s_{1/2}$    &&   13.7\,  ($-$2.7) & 0.2 & [$+$$+$] & \multirow{2}{*}{$\Big\}$\, \ 13.8}  
                                                                           & \multirow{2}{*}{0.7$\pm$1.0}  \\
$1^+_2$  &&  $3/2^+$, $1/2^+$  &$0s_{1/2}$   &&   16.3\,  ($-$0.18) & 0.2 & [$+$$+$] &          &        \\
\end{tabular}
\end{ruledtabular}
\end{table*}
%%%%%%%%%%%%%%%%%%%%%%%%%%%%%%%%%%%%%%%%%%%%%%%%%%%%%%%

%%%%%%%%%%%%%%%%%%%%%%%%%%%%%%%%%%%%%%%%%%%%%%%%%%%%%%%%%%%%%%%%%%%%%%%%%%%%%%%%%
%\Figuretable{FIG. 5}
% Figure 5
\begin{figure}[tb]
\begin{center}
  \includegraphics[width=0.8\linewidth]{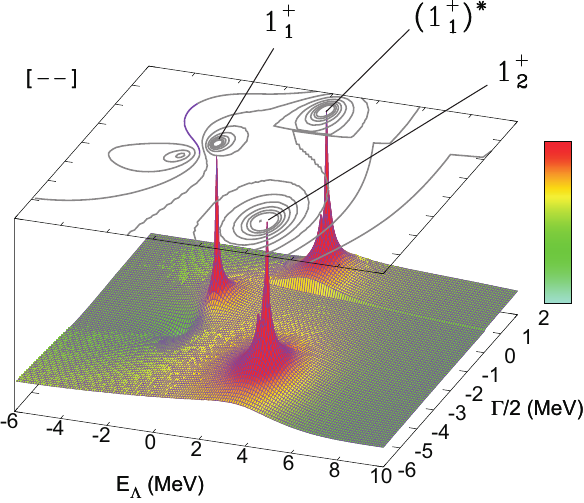}
\end{center}
\caption{\label{fig:5}
Pole positions for the $J_B=$ $1^+$ states 
in the complex $E$ plane \cite{Pearce89,Harada17b},
using a two-channel coupled system with 
${^5{\rm Li}(3/2^-)}\otimes (0p_{3/2})_\Lambda$ and 
${^5{\rm Li}(3/2^-)}\otimes (0p_{1/2})_\Lambda$ channels 
within the ($\alpha$-$p$)-$\Lambda$ model space.
Poles for $1^+_1$ and $1^+_2$ reside at ${\cal E}_\Lambda = -0.45 -i 0.98$~MeV 
and ${\cal E}_\Lambda= 3.1 -i 2.9$~MeV on the [$-$$-$] Riemann sheet, respectively.
A pole for $(1^+_1)^*$ is a partner pole associated with 
the $1^+_1$ pole. 
}
\end{figure}
%%%%%%%%%%%%%%%%%%%%%%%%%%%%%%%%%%%%%%%%%%%%%%%%%%%%%%%%%%%%%%%%%%%%%%%%%%%%%%%%%%

%%%%%%%%%%%%%%%%%%%%%%%%%%%%%%%%%%%%%%%%%%%%%%%%%%%%%%%%%%%%%%%%%%%%%%%%%%%%%%%%%
%\Figuretable{FIG. 6}
% Figure 6
\begin{figure}[tb]
\begin{center}
  \includegraphics[width=0.8\linewidth]{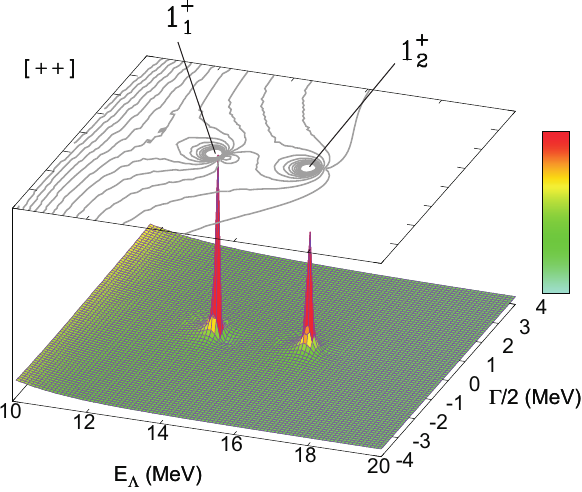}
\end{center}
\caption{\label{fig:6}
Pole positions for the $J_B=$ $1^+$ states 
in the complex $E$ plane \cite{Pearce89,Harada17b}, 
using a two-channel coupled system with 
${^5{\rm Li}(3/2^+)}\otimes (0s_{1/2})_\Lambda$ and 
${^5{\rm Li}(1/2^+)}\otimes (0s_{1/2})_\Lambda$ channels 
within the ($^3{\rm He}$-$d$)-$\Lambda$ model space.
Poles for $1^+_1$ and $1^+_2$ reside at 
${\cal E}_\Lambda=  13.7 -i 0.1$~MeV and 
${\cal E}_\Lambda=  16.3 -i 0.1$~MeV 
on the [$+$$+$] Riemann sheet, respectively.
}
\end{figure}
%%%%%%%%%%%%%%%%%%%%%%%%%%%%%%%%%%%%%%%%%%%%%%%%%%%%%%%%%%%%%%%%%%%%%%%%%%%%%%%%%%

We discuss the eigenstates of $^6_\Lambda$Li obtained 
by solving the CC equations with the $\Lambda$-nucleus potentials 
given in Eq.~(\ref{eqn:e9}) for ${^5{\rm Li}(J_C)}$-$\Lambda$ systems 
within the ($\alpha$-$p$)-$\Lambda$ and ($^3{\rm He}$-$d$)-$\Lambda$ model spaces. 
Table~\ref{tab:table2} presents the calculated energies and widths 
of the $\Lambda$-bound and resonance states with 
$[{^5{\rm Li}(J_C)}\otimes{j_\Lambda}]_{J_B}$ in $^6_\Lambda$Li. 

For $(\alpha\mbox{-}p)\mbox{-}\Lambda$ configurations, 
the $1^-$ ground state (g.s.) is found at $E_\Lambda=$ $-$4.51 MeV, 
and the $2^-$ excited state (exc.) at $E_\Lambda=$ $-$4.26 MeV, 
both lying below the ${^5{\rm Li}({\rm g.s.})}+\Lambda$ threshold. 
These $1^-$(g.s.) and $2^-$(exc.) bound states are unstable 
due to their strong interaction decay into $^5_\Lambda{\rm He}+p$, 
located at $E_\Lambda=$ $-$5.08 MeV.
The spin-spin coupling between $^5{\rm Li}(3/2^-)$ and 
$^5{\rm Li}(1/2^-)$ with $\Lambda$ for the $1^-$(g.s.)
is not so significant 
because the ${\bm I}\cdot{\bm S}$ coupling potential is small, 
as mentioned in Sec.~\ref{potential}.

The nature of a bound or resonance state is often determined by 
the pole position of 
the $S$ matrix in an $n$-channel coupled system, 
where the $2^n$ Riemann sheets are characterized by $n$ signs 
of $[{\rm Im}k_{1}\,{\rm Im}k_{2}\,\cdots\,{\rm Im}k_{n}]$ 
with the c.m.~momentum 
$k_{c}=(k_1, k_2, \cdots, k_n)$ in the complex $k$ plane. 
Thus, the complex energy ${\cal E}_\Lambda$ can be expressed as
\begin{eqnarray}
{\cal E}_\Lambda = \frac{k_c^2}{2\mu_c}+\Delta_{c}= E_\Lambda - i \frac{1}{2}\Gamma,
\label{eqn:e15}
\end{eqnarray}
where $\Delta_c$ denotes the $c$ channel threshold energy 
measured from the $^5{\rm Li}({\rm g.s.})+\Lambda$ threshold. 
For $J_B=$ $1^+$, we find the $p$-wave poles of the $S$-matrix 
in the $n=2$ coupled system that mainly consists of 
$^5{\rm Li}(3/2^-)\otimes (0p_{3/2})_\Lambda$ and 
$^5{\rm Li}(3/2^-)\otimes (0p_{1/2})_\Lambda$ 
configurations on the [$-$$+$] and [$-$$-$] Riemann sheets. 
In Fig.~\ref{fig:5}, we illustrate the pole structure for $J_B=$ $1^+$ 
near the $^5{\rm Li}({\rm g.s.})+\Lambda$ threshold on 
the [$-$$-$] sheet in the complex $E$ plane \cite{Pearce89,Harada17b}.
The $1^+_1$ pole is located at ${\cal E}_\Lambda = -0.45 -i 0.98$~MeV 
on the [$-$$-$] sheet, which is identified as a $0p_\Lambda$ (decaying) resonance. 
Because the $\Lambda$-nucleus potential has 
an imaginary part of $-0.6$ MeV due to the unstable $^5{\rm Li}({\rm g.s.})$ core, 
this pole shifts closer to the threshold and behaves like a virtual state. 
A continuum state for $1^+$ shows an enhancement above the 
threshold (2--3 MeV) by the nearby $1^+_1$ pole. 
As a result, the shape of the spectrum changes near the threshold \cite{Morimatsu94}.
On the other hand, the $(1^+_1)^*$ pole is a partner of the $1^+_1$ pole, 
which corresponds to a $0p_\Lambda$ (capturing or UBS \cite{Morimatsu94}) resonance, originating from a 
$p$-wave conjugate pole associated with the $1^+_1$ pole, 
not observable in the spectrum due to its large distance 
from the physical axis \cite{Morimatsu94}. 

Moreover, we find the $1^+_2$ pole at ${\cal E}_\Lambda= 3.1 -i 2.9$~MeV 
on the [$-$$-$] sheet, which is identified as an excited $1p_\Lambda$ resonance pole 
in the $S$ matrix trajectory obtained using complex potentials, 
as discussed by Dabrowski \cite{Dabrowski97}. 
However, this pole shifts far from the physical axis 
due to the imaginary potential, resulting in a broad width of $\Gamma \simeq$ 5.6 MeV 
in the $\Lambda$ continuum. 
These results suggest that the $J_B=1^+$ state above the 
$^5{\rm Li}({\rm g.s.})+\Lambda$ threshold does not form a broad resonance 
but rather constitutes a continuum state influenced by nearby poles.
This state has perhaps been identified as a substitutional $(0p,~0p^{-1})_{\Lambda n}$ state 
within shell models and cluster models \cite{Majling80,Motoba83,Motoba85}.
Therefore, we conclude that the broad peak at 3.8 MeV shown in Fig.~\ref{fig:1} corresponds to 
a continuum state influenced by nearby poles.

For $({^3{\rm He}}\mbox{-}d)\mbox{-}\Lambda$ configurations, 
two poles, $1^+_1$ and $1^+_2$, exist at 
${\cal E}_\Lambda=$ $13.7-i0.1$ MeV and $16.3-i0.1$ MeV, respectively, 
close to the physical axis, as shown in Table~\ref{tab:table2}. 
Owing to the weak coupling between 
the $({^3{\rm He}}\mbox{-}d)\mbox{-}\Lambda$ and 
$(\alpha\mbox{-}p)\mbox{-}\Lambda$ channels \cite{Majling80}, 
the $1^+_1$ and $1^+_2$ states may approximately behave as $\Lambda$ quasi-bound states, 
which have $^5{\rm Li}(3/2^+)\otimes (0s_{1/2})_\Lambda$ and 
$^5{\rm Li}(1/2^+)\otimes (0s_{1/2})_\Lambda$ components, 
on the [$+$$+$] Riemann sheet below the ${^3{\rm He}}+d+\Lambda$ threshold. 
The $1^+_1$ state is located at $E_\Lambda=$ 13.7 MeV, 
corresponding to $E_\Lambda - E_{\rm th}= -2.7$ MeV 
near the ${^3{\rm He}}+d+\Lambda$ threshold, where $E_{\rm th}=$ 16.39 MeV represents 
the ${^3{\rm He}}+d$ threshold energy 
measured from the ${^5{\rm Li}({\rm g.s.})}$, as shown in Fig.~\ref{fig:2}. 
This state is identified as a $\Lambda$ quasi-bound state having the rms radius 
of $\langle r^2_\Lambda \rangle^{1/2}=$ 3.2 fm, where 
the mixing probability of ${^5{\rm Li}(1/2^+)}$ into ${^5{\rm Li}(3/2^+)}$ for $1^+_1$ 
amounts to 6.8\% due to the ${\bm I}\cdot{\bm S}$ coupling, 
which is comparable to 6\% obtained in the 
${^3{\rm He}}+d+\Lambda$ cluster model~\cite{Ohkura93}.
The $1^+_2$ state is located at $E_\Lambda=$ 16.3 MeV, corresponding to
$E_\Lambda - E_{\rm th}= -0.18$ MeV, 
and is identified as a loosely $\Lambda$ quasi-bound state with 
a large distance of $\langle r_\Lambda^2 \rangle^{1/2}=$ 6.1 fm.
Figure~\ref{fig:6} illustrates the pole structures for $J_B=$ $1^+$ 
near the ${^3{\rm He}}+d+\Lambda$ threshold 
on the [$+$$+$] Riemann sheet \cite{Pearce89,Harada17b}.

\subsection{Comparison with the experimental data}

%%%%%%%%%%%%%%%%%%%%%%%%%%%%%%%%%%%%%%%%%%%%%%%%%%%%%%%
%\Figuretable{FIG. 7}
% Figure 7
\begin{figure}[t]
\begin{center}
  \includegraphics[width=1.0\linewidth]{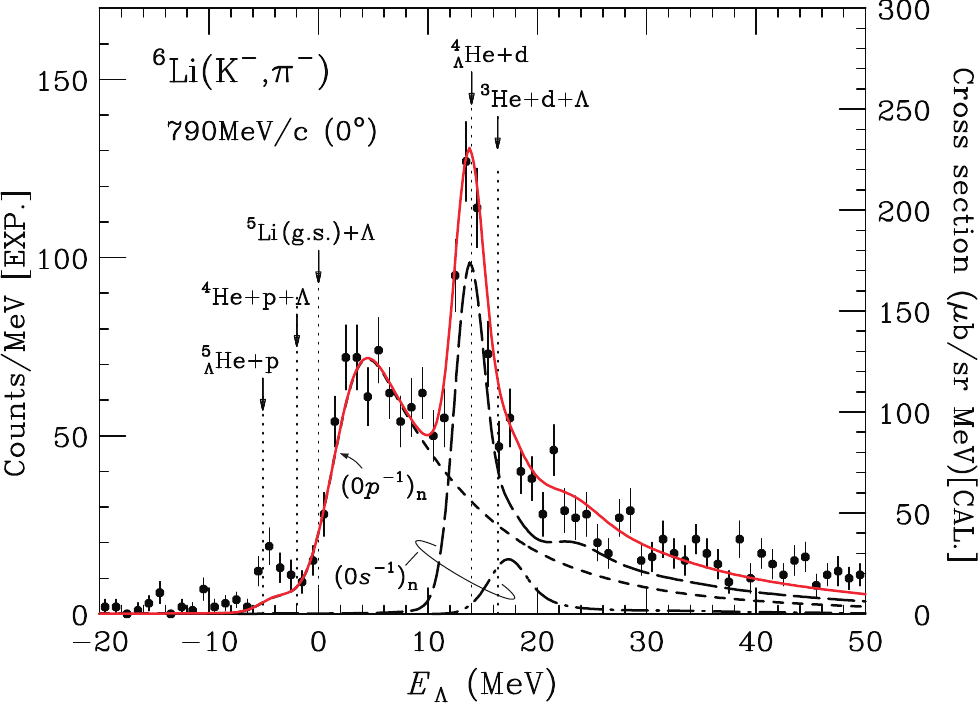}
\end{center}
\caption{\label{fig:7}
Calculated inclusive $\pi^-$ spectrum for $\Lambda$ production via the 
$^6$Li($K^-$,~$\pi^-$) reaction at $p_{K^-}=$ 790 MeV/$c$ 
and $\theta_{\rm lab}=$ 0$^\circ$, together 
with the experimental data \cite{Bertini81a}, 
as a function of $E_\Lambda$ measured relative to 
the ${^5{\rm Li}({\rm g.s.})}+\Lambda$ threshold.
Solid, dashed and long-dashed curves denote neutron-hole contributions of the total, 
$0p^{-1}$, and $0s^{-1}$, respectively, 
using $\overline{f}_{K^-n \to \pi^-\Lambda}$ obtained by EOFA \cite{Harada22}.
A dot-dashed curve denotes an additional contribution formed 
by a direct $^4_\Lambda{\rm He}+d$ reaction.
These spectra are considered by a detector resolution of 4 MeV FWHM. 
}
\end{figure}
%%%%%%%%%%%%%%%%%%%%%%%%%%%%%%%%%%%%%%%%%%%%%%%%%%%%%%%

We analyze the $\Lambda$ production spectrum for 
the ($K^-$,~$\pi^-$) reaction on a $^6{\rm Li}$ target 
in the DWIA with the Fermi-averaged $K^-n\to\pi^-\Lambda$ amplitudes 
obtained by EOFA. 
Because a nuclear ($K^-$,~$\pi^-$) reaction is exothermic, 
this reaction satisfies under a near-recoilless condition 
by controlling a momentum transfer \cite{Feshbach66,Kerman71}. 
For $K^-$ beams at $p_{K^-}=$ 700--800 MeV/$c$ and 
$\theta_{\rm lab}=$ 0$^\circ$, the momentum transfer is 
$q =$ 50--80 MeV/$c$ and $q \simeq$ 51 MeV/$c$ 
at the $\Lambda$ production threshold, 
leading to that the angular-momentum transfer $\Delta L =$ 0 dominates. 
For a $^6$Li($1^+$; g.s.) target nucleus, therefore,  
the substitutional states having $J_B=$ 1$^+$ are expected 
to be predominantly populated in $\Lambda$ hypernuclear 
states \cite{Feshbach66,Kerman71,Povh76,Bruckner76,Bruckner78}. 

Figure~\ref{fig:7} displays the calculated inclusive $\pi^-$ spectrum for 
$\Lambda$ production via the $^6$Li($K^-$,~$\pi^-$) reaction 
at $p_{K^-}=$ 790 MeV/$c$ and $\theta_{\rm lab}=$ 0$^\circ$, 
together with the experimental data \cite{Bertini81a}. 
The calculated spectrum considers a detector resolution of 4 MeV FWHM. 
The shape of the calculated spectrum agrees well with the data 
across the entire $\Lambda$ production region, except for a small peak 
in the $\Lambda$ bound region.
Furthermore, it should be noted that the absolute values of 
the magnitude of the calculated spectrum are 
approximately half smaller than that suggested 
by the experimental data \cite{Povh76,Bruckner76,Bruckner78}. 
This situation suggests a similar trend to that theoretical values 
obtained using EOFA for the angular distributions can reproduce 
the data of the $^{12}$C($K^-$,~$\pi^-$)$^{12}_\Lambda$C reaction 
measured by the BNL experiments \cite{Chrien79},  
as discussed in a previous work \cite{Harada22}. 
Therefore, we show that the calculated spectra predict the data 
of the $^{6}$Li($K^-$,~$\pi^-$)$^{6}_\Lambda$Li reaction.

\section{Discussion}

\subsection{Partial spin components in the spectrum}

Figure~\ref{fig:8} shows the contributions from neutron 
$0p^{-1}$ and $0s^{-1}$ states to the 
inclusive $\pi^-$ spectrum shown in Fig.~\ref{fig:7}.  
We estimate the partial spin $J^p$ contributions of 
the $\Lambda$ production for the corresponding ($\alpha$-$p$)-$\Lambda$ and 
($^3{\rm He}$-$d$)-$\Lambda$ configurations, 
as discussed in the following subsections. 

\subsubsection{
$(\alpha\mbox{-}p)\mbox{-}\Lambda$ spectra}

%%%%%%%%%%%%%%%%%%%%%%%%%%%%%%%%%%%%%%%%%%%%%%%%%%%%%%%
%\Figuretable{FIG. 8}
% Figure 8
\begin{figure}[tb]
\begin{center}
  \includegraphics[width=0.9\linewidth]{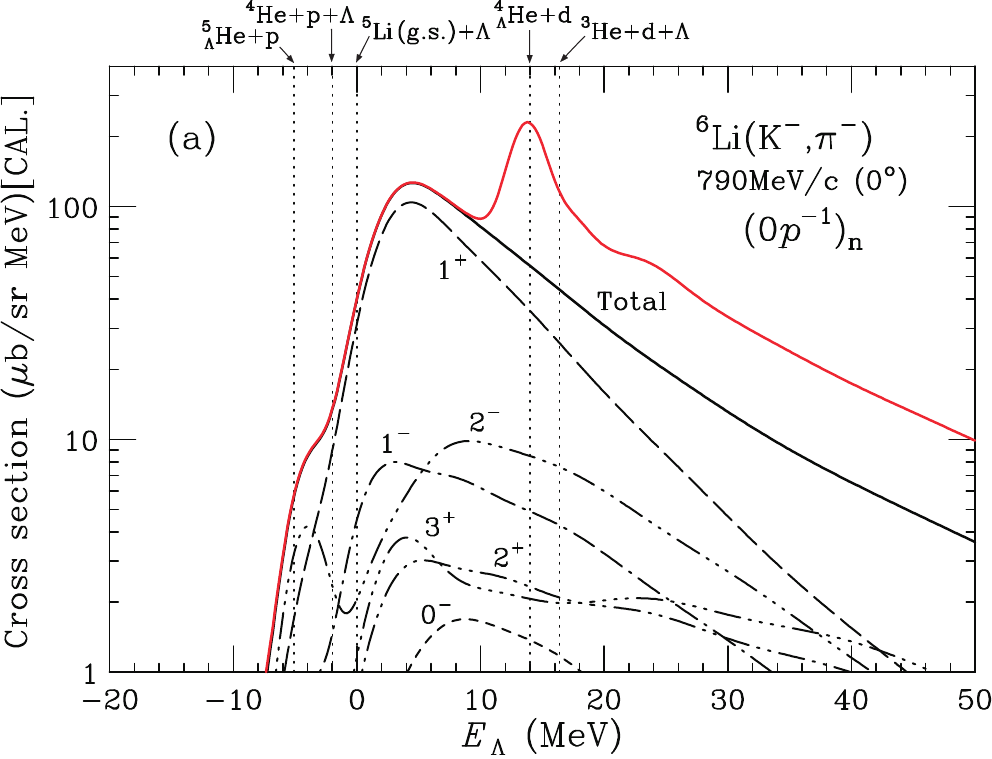}
  \includegraphics[width=0.9\linewidth]{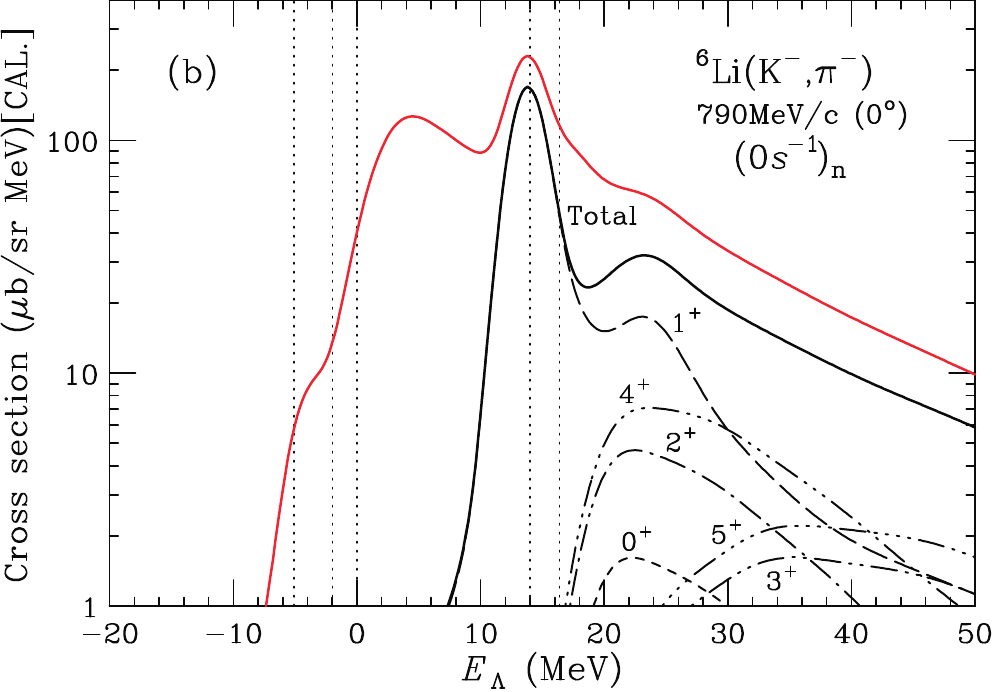}
\end{center}
\caption{\label{fig:8}
Partial spin $J^P$ contributions of $^6_\Lambda$Li hypernuclear states 
from neutron (a) $0p^{-1}$ and (b) $0s^{-1}$ components, 
to the  $\Lambda$ production spectra in $^6$Li($K^-$,~$\pi^-$) reactions 
at 790 MeV/$c$ and $\theta_{\rm lab}=$ 0$^\circ$. 
These components are obtained in ($\alpha$-$p$)-$\Lambda$ and 
($^3{\rm He}$-$d$)-$\Lambda$ configurations, respectively. 
The spectra are considered with a detector resolution of 4 MeV FWHM. 
}
\end{figure}
%%%%%%%%%%%%%%%%%%%%%%%%%%%%%%%%%%%%%%%%%%%%%%%%%%%%%%%

For low-lying states with ($\alpha$-$p$)-$\Lambda$ configurations, 
the cross section of the $J_B=$ $2^-$(exc.) state at 
$E_\Lambda=$ $-$4.26 MeV dominates below the 
${^5{\rm Li}({\rm g.s.})}+\Lambda$ threshold, as shown in Fig.~\ref{fig:8}(a). 
In contrast, the cross section of the $1^-$(g.s.) state at $E_\Lambda=$ $-$4.51 MeV 
is negligible in the $\Lambda$ bound region 
due to the suppressed $\Lambda$ production amplitude, resulting from 
an admixture of both $(0p_{3/2},~0s_{1/2}^{-1})_{\Lambda n}$ and 
$(0p_{1/2},~0s_{1/2}^{-1})_{\Lambda n}$ components.
This situation may correspond to an admixture of $S_>$ and $S_<$ states 
as suggested by cluster-model calculations \cite{Motoba83,Motoba85}.
Since the $2^-$(exc.) component dominates in the $\Lambda$ bound region, 
the small peak observed at $E_\Lambda \simeq$ $-$4.50 MeV 
in the data  is interpreted as the $2^-$(exc.) state. 
This result is consistent with that obtained in cluster-model 
calculations \cite{Motoba83,Motoba85}.
However, our calculations do not well reproduce the pronounced small 
peak, assuming a detector resolution of 4 MeV FWHM.

The cross section for the $1^+$(exc.) state associated 
with ${^5{\rm Li}({\rm g.s.})}\otimes 0p_\Lambda$, 
which corresponds to the $(0p,~0p^{-1})_{\Lambda n}$ substitutional component, 
is predominantly populated. 
This state manifests as a broad peak at $E_\Lambda\simeq$ 5.0 MeV 
above the ${^5{\rm Li}({\rm g.s.})}+\Lambda$ threshold, as shown in Fig.~\ref{fig:7}.
The poles for $1^+$(exc.) are located on the unphysical sheet [$--$]
at ${\cal E}_\Lambda = -0.45 -i 0.98$~MeV close to the threshold, and 
at ${\cal E}_\Lambda = 3.1 -i 2.9$~MeV 
which exhibits a broad width and lies far from the physical axis, 
as discussed in Sec.~\ref{levels}.
This suggests that the broad peak of the $1^+$ state 
stems from $\Lambda$ continuum states influenced by the nearby poles.
This result highlights the crucial role of continuum effects 
in $\Lambda$ production, as observed in the $\pi^-$ spectra \cite{Morimatsu94}, 
indicating that bound-state approximations 
are insufficient for accurately describing
continuum spectra \cite{Motoba83,Motoba85,Ohkura93}.
Therefore, a rigorous approach, such as the Green's function 
method \cite{Morimatsu94}, 
is crucial for a precise analysis of $\Lambda$ production spectra 
in nuclear ($K^-$,~$\pi^-$) reactions.
This approach facilitates a comprehensive description 
of $\Lambda$-nucleus dynamics,
including both bound and continuum states.

\subsubsection{
$({^3{\rm He}\mbox{-}d})\mbox{-}\Lambda$ spectra}

%%%%%%%%%%%%%%%%%%%%%%%%%%%%%%%%%%%%%%%%%%%%%%%%%%%%%%%
%\Figuretable{FIG. 9}
% Figure 9
\begin{figure}[t]
\begin{center}
  \includegraphics[width=1.0\linewidth]{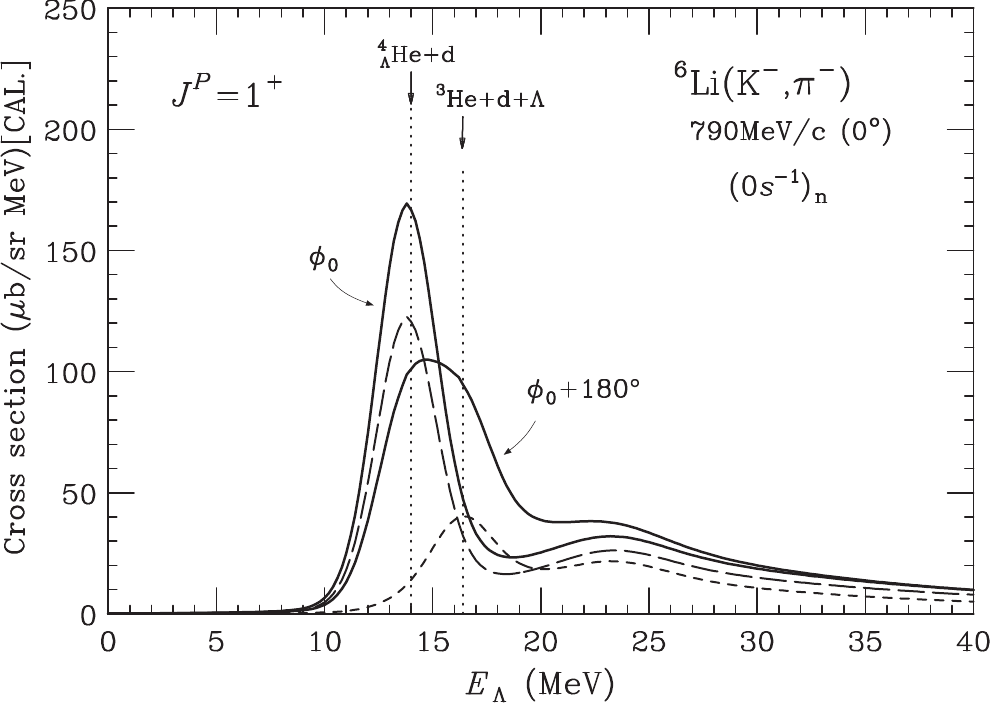}
\end{center}
\caption{\label{fig:9}
Interference effects of the $\Lambda$ production for $J^P=$ $1^+$ near the 
${^3{\rm He}}+d+\Lambda$ threshold on the $^6$Li($K^-$,~$\pi^-$) 
reaction at $p_{K^-}=$ 790 MeV/$c$ and $\theta_{\rm lab}=$ 0$^\circ$.
The solid curves denote the total spectra changing the relative phases of 
$\phi=$  $\phi_0$ and $\phi_0+180^\circ$ 
between $T_{3}$ and $T_{1}$ amplitudes. 
The long-dashed and dashed curves denote the contributions 
of $|T_3|^2$ and $|T_1|^2$, respectively. 
The spectra are considered with a detector resolution of 4 MeV FWHM. 
}
\end{figure}
%%%%%%%%%%%%%%%%%%%%%%%%%%%%%%%%%%%%%%%%%%%%%%%%%%%%%%%

For high-lying states with ($^3{\rm He}\mbox{-}d$)-$\Lambda$ configurations, 
the cross section for the $J_B=$ $1^+$ state 
is predominantly populated, which exhibits a narrow peak 
($\Gamma^{\rm exp} \simeq$ 0.7$\pm$1.0 MeV) near the ${^4_\Lambda{\rm He}}+d$ threshold, 
as shown in Fig.~\ref{fig:8}(b). 
Other higher-spin $J^P$ states are rarely populated. 
This component with $J^P=$ $1^+$ originates from 
${^5{\rm Li}}(3/2^+)\otimes(0s_{1/2})_\Lambda$
and ${^5{\rm Li}}(1/2^+)\otimes(0s_{1/2})_\Lambda$ configurations, corresponding 
to nearly pure $(0s,~0s^{-1})_{\Lambda n}$ 
substitutional states due to the small momentum transfer $q \simeq$ 50--80 MeV/$c$. 
To examine the contributions of the $1^+_1$ and $1^+_2$ states, whose poles reside 
at $E_\Lambda=$ 13.7 MeV and 16.3 MeV in Table~\ref{tab:table2}, 
respectively, we consider the double-differential cross section for $1^+$(exc.) 
in Eq.~(\ref{eqn:e1}), which can be expressed as
\begin{eqnarray}
\frac{d^2\sigma(1^+)}{{d\Omega}{dE}}
 & \propto & |T_{3}|^2+|T_{1}|^2 + 2{\rm Re}(T_{3}^* T_{1}),
\label{eqn:e14}
\end{eqnarray}
where $T_{3}$ and $T_{1}$ represent the $\Lambda$ production amplitudes 
for ${^5{\rm Li}(3/2^+)}\otimes(0s_{1/2})_\Lambda$ 
and ${^5{\rm Li}(1/2^+)}\otimes(0s_{1/2})_\Lambda$ components, respectively, 
and the relative phase between them is given by $\phi_0= {\rm Arg}(T_{3}^* T_{1})$. 
In Fig.~\ref{fig:9}, we present several spectra of the $1^+$ cross section 
by artificially modifying the relative phases to $\phi =$ $\phi_0$ 
and $\phi_0+180^\circ$ in the interference term 
$2{\rm Re}(T_{3}^* T_{1})$ in Eq.~(\ref{eqn:e14}), 
along with the individual contributions of $|T_{3}|^2$ and $|T_{1}|^2$. 
The interference term modifies the spectrum, 
resulting in a narrower peak shape near the ${^3{\rm He}}+d+\Lambda$ threshold, 
although both $1^+_1$ and $1^+_2$ states contribute to the $\Lambda$ production. 
Additionally, we observe a broad shoulder in $1^+$ around $E_\Lambda=$ 24 MeV, 
as shown in Fig.~\ref{fig:8}(b), which originates from 
the ${^5{\rm Li}}(3/2^+,1/2^+)\otimes (0d_{5/2,3/2})_\Lambda$ 
configurations obtained from the ($^3{\rm He}\mbox{-}d$)-$\Lambda$ 
cluster potentials.

\subsubsection{
${^4_\Lambda{\rm He}}+d$ spectra}

It is worthwhile to note that light nuclear targets 
such as $^6$Li have played an essential role as a doorway 
for producing various hyperfragments,  
including $^5_\Lambda$He, $^4_\Lambda$He, $^4_\Lambda$H, and $^3_\Lambda$H, 
via the $(K^-$,~$\pi^-$) reactions \cite{Majling80}. 
To examine the contribution of a ${^4_\Lambda{\rm He}}+d$ process, 
which corresponds to the direct production of a 
$^4_\Lambda$He hyperfragment \cite{Tamura89}, 
we estimate the cross section for ${^4_\Lambda{\rm He}}+d$ production 
via the reaction $K^-+{^6{\rm Li}}\to\pi^-+{^4_\Lambda{\rm He}}+d$ 
at $p_{K^-}=$ 790 MeV/$c$ and $\theta_{\rm lab}=$ 0$^\circ$, 
using the form factor $F_\alpha(q)$ for a $^4$He target 
within a DWIA calculation \cite{Harada19,Harada21}.  
In Fig.~\ref{fig:7}, we have presented the $\pi^-$ spectra 
at $p_{K^-}=$ 790 MeV/$c$ and $\theta_{\rm lab}=$ 0$^\circ$, 
including contributions from the ${^4_\Lambda{\rm He}}+d$ component, 
along with those from the $(\alpha\mbox{-}p)\mbox{-}\Lambda$ 
and $(^3{\rm He}\mbox{-}d)\mbox{-}\Lambda$ components. 
The contribution of 
the ${^4_\Lambda{\rm He}}+d$ component in the $(K^-$,~$\pi^-$) 
reaction at $p_{K^-}=$ 790 MeV/$c$ and $\theta_{\rm lab}=$ 0$^\circ$ 
is not so significant, as the continuum states associated 
with ${^4_\Lambda{\rm He}}+d$ are naturally suppressed 
due to the small momentum transfer of $q \simeq$ 50--80 MeV/$c$ 
in this substitutional reaction.

\subsection{\boldmath
Effects of the three-body $YNN$ force and $\Sigma$ admixture}

The high-lying excited states of $^6_\Lambda$Li appear to be promising  
candidates for investigating the three-body $YNN$ force involving  
$\Lambda NN$-$\Sigma NN$ coupling induced by the $\Lambda N$-$\Sigma N$  
interaction, rather than the Fujita-Miyazawa-type $\Lambda NN$ force,  
which is expected to have only a minor impact on the energies in free  
space.  
In the (${^3{\rm He}}$-$d$)-$\Lambda$ configuration, the  
${^3{\rm He}}+\Lambda$ subsystem is coupled with  
${^3{\rm H}}+\Sigma^+$ and ${^3{\rm He}}+\Sigma^0$ in $^4_\Lambda$He via  
the $\Lambda NN$-$\Sigma NN$ interaction caused by coherent  
$\Lambda$-$\Sigma$ coupling~\cite{Harada00,Akaishi00}, and the $d+\Lambda$  
subsystem is coupled with $\{nn\}\Sigma^+$, $\{pn\}\Sigma^0$, and  
$\{pp\}\Sigma^-$ in $^3_\Lambda$H~\cite{Harada14}.  
Therefore, the three-body $YNN$ force is expected to play a crucial role  
in determining the fine structure of the excited  
(${^3{\rm He}}$-$d$)-$\Lambda$ states, depending on the spin-isospin  
states in $^5{\rm Li}(J_C)$-$\Lambda/\Sigma$.  
However, the present analysis cannot resolve the energy shifts of the  
fine $\Lambda$ states, as the experimental spectra were measured with a  
resolution of approximately 4 MeV FWHM~\cite{Bertini81a}.  
To clarify the effects of the $YNN$ force in this nucleus, high-resolution  
measurements at dedicated beamlines are required.  

Moreover, $\Lambda$-$\Sigma$ coupling induces a $\Sigma$ admixture in  
the wave functions of $\Lambda$ hypernuclear states~\cite{Umeya09}.  
These $\Sigma$ components can also be populated through direct processes  
such as $K^-p \to \pi^-\Sigma^+$ and $K^-n \to \pi^-\Sigma^0$, which  
provide direct access to the $\Sigma$ admixture in the nucleus~\cite{Harada09}.  
However, this effect is expected to be negligible in the nuclear  
($K^-$,~$\pi^-$) reaction at $p_{K^-} = 790$ MeV/$c$, as the  
Fermi-averaged amplitudes  
$\overline{f}_{K^-p \to \pi^-\Sigma^+}$ and  
$\overline{f}_{K^-n \to \pi^-\Sigma^0}$ are much smaller than  
$\overline{f}_{K^-n \to \pi^-\Lambda}$.

\subsection{Comparison with other calculations}

Majling {\it et al}.~\cite{Majling80} discussed that the peak at 13.8 MeV 
in the $^6$Li($K^-$,~$\pi^-$) data indicates $\Lambda$ production 
via $(0s, 0s^{-1})_{\Lambda n}$ with a narrow width 
of $\Gamma \simeq$ 0.7 $\pm$ 1.0 MeV, 
due to the small admixture between wave functions with 
$[f]=$ [411] and [321] in spatial symmetry.
According to Ref.~\cite{Majling80}, 
we approximately neglect the couplings between the ($\alpha\mbox{-}p$)-$\Lambda$ 
and ($^3{\rm He}\mbox{-}d$)-$\Lambda$ channels in the CC equation.
This assumption is further supported by the small imaginary part of 
the optical potential for $t + d$ scattering \cite{Neudatchin91,Tilley02} 
and the reduction of events near the peak at 13.8 MeV 
in the spectrum for proton coincidence via nonmesonic weak decay 
of $^5_\Lambda$He \cite{Szymanski86}. 
Therefore, we conclude that this approximation is valid in our calculations.  

Auerbach and Van Giai \cite{Auerbach80} attempted to calculate continuum spectra 
within a single-particle $\Lambda$ description 
using a $\Lambda N^{-1}$ basis for $^6$Li, $^7$Li, and $^9$Be targets, 
revealing the fundamental features of continuum effects 
in nuclear ($K^-$,~$\pi^-$) reactions. 
They indicated that the peak at 3.5 MeV in $^6_\Lambda$Li does not correspond 
to a resonance state but rather to a continuum state 
above the ${^5{\rm Li}({\rm g.s.})}+\Lambda$ threshold.
However, their calculated spectra fail to accurately reproduce 
the experimental data across the entire energy range 
due to the omission of spin structure in the 
${^5{\rm Li}}(J_C)\otimes \Lambda$ configuration. 
When evaluating the continuum states in the spectra, 
it is essential to incorporate the spin structure of $^6_\Lambda$Li, 
as discussed in Sec.~\ref{Results}. 

Motoba {\sl et al}.~\cite{Motoba83,Motoba85} 
investigated the structures and production cross sections 
in the ($K^-$,~$\pi^-$) reaction on a $^6$Li target using a microscopic 
$\alpha+p+\Lambda$ cluster model.
They analyzed the spin structure 
in $^6_\Lambda$Li, demonstrating that the 
${\bm \sigma}_N\cdot{\bm \sigma}_\Lambda$ term 
plays a crucial role in determining the cross sections. 
Their findings indicate that 
the production cross section for the $2^-$(exc.) state is 
larger than that for the $1^-$(g.s.) state 
because the spin $S_>$ and  $S_<$ components are significantly mixed 
only in the $1^-$(g.s.) state. 
The peak at 3.8 MeV has been identified as a broad $1^+$ resonance, 
which does not have a simple shell-model substitutional structure 
such as $(p_{3/2},~p_{3/2}^{-1})_{\Lambda n}$. 
Instead, it exhibits a strong admixture of 
$^5{\rm Li}(3/2^-)\otimes (0p_{3/2})_\Lambda$ and 
$^5{\rm Li}(3/2^-)\otimes (0p_{1/2})_\Lambda$ 
within the cluster-model framework. 
However, our results suggest that no broad resonance peak exists for 
the $1^+({\rm exc.})$ state, despite the significant role of the spin structure 
in describing the spectrum data.

Ohkura {\sl et al}.~\cite{Ohkura93} extensively studied the structure 
and production mechanisms of 
a high-lying excited state of $^6_\Lambda$Li using a microscopic 
${^3{\rm He}} + d + \Lambda$ cluster model. 
Their calculated energies and widths of the 
$1^+_1$ and $1^+_2$ excited states successfully 
explain the prominent peak at 13.8 MeV observed 
in the $^6$Li($K^-$,~$\pi^-$) reaction data. 
Our results appear to be similar to theirs. 
However, since their calculations are based on a bound-state approximation, 
they fail to describe the continuum $\Lambda$ states accurately 
above the ${^3{\rm He}}+d+\Lambda$ threshold.

Therefore, to extract valuable insights into the structure and production mechanisms 
of $\Lambda$ hypernuclei, 
it is crucial to compute the $\Lambda$ production spectra 
while explicitly incorporating the spin structure of $\Lambda$ bound, resonance, 
and continuum states in $^6_\Lambda$Li, as discussed in Sec.~\ref{Results}.

\subsection{Future outlook}

It is important to discuss potential improvements in the theoretical  
framework of our calculations to achieve our goals. One key aspect of  
such improvements is the consideration of self-consistency in  
describing the structure of light $\Lambda$ hypernuclei, particularly  
for highly excited states near the nuclear core breakup thresholds,  
at which the $\alpha$ cluster breaks into $3N+N$. This consideration is  
crucial because the shrinkage effects induced by the $\Lambda$  
hyperon~\cite{Hiyama96} can enhance the nuclear core density, requiring  
more accurate microscopic descriptions of the core states.  

The fine structure of high-lying excited states in $^6_\Lambda$Li is  
important for examining the effects of the three-body $YNN$ force,  
which involves $\Lambda NN$-$\Sigma NN$ coupling induced by the  
$\Lambda N$-$\Sigma N$ interaction in the $\Lambda$ production mechanism. 
To achieve a comprehensive understanding of such systems, it is also  
essential to adopt a unified treatment for describing the production  
reactions of $\Lambda$ hypernuclei. The reaction channel  
(${^3{\rm He}}$-$d$)-$\Lambda$ involves rearrangement channels  
corresponding to (${^3{\rm He}}$-$\Lambda$)-$d$. Thus, a consistent  
theoretical approach to these rearrangement-channel reactions is  
essential. Moreover, a unified approach to the $\Lambda$ production  
reaction in both the ($\alpha$-$p$)-$\Lambda$ and  
(${^3{\rm He}}$-$d$)-$\Lambda$ channels is required to perform a  
fully coupled calculation that accurately describes the reaction  
dynamics.

\section{Summary and Conclusion}
\label{Conclusion}

We have conducted a theoretical investigation of the $\Lambda$ production spectra 
via the ($K^-$,~$\pi^-$) reaction on a $^6$Li target, using DWIA 
with a Fermi-averaged $K^-n \to \pi^- \Lambda$ amplitude obtained through EOFA.
We have calculated the $\Lambda$ bound, resonance, and continuum states 
in the Green's function method for the ${^5{\rm Li}}\mbox{-}\Lambda$ system 
by solving the CC equations with the $\Lambda$ folding-model potentials 
based on $\alpha\mbox{-}p$ and ${^3{\rm He}}\mbox{-}d$ cluster wave functions. 
We have analyzed the shape and magnitude of the spectra by comparing them 
with experimental data from the $^6$Li($K^-$,~$\pi^-$) reaction. 
Results are summarized as follows:

\begin{itemize}
\item[(1)] 
The calculated spectrum for the ($K^-$,~$\pi^-$) reaction at 
$p_{K^-}=$ 790 MeV/$c$ (0$^\circ$) 
agrees well with the experimental data, incorporating substitutional 
$(0p,~0p^{-1})_{\Lambda n}$ and $(0s,~0s^{-1})_{\Lambda n}$ configurations. 
It is crucial to consider the spin structure of ${^6_\Lambda{\rm Li}}$ 
to accurately evaluate the spectra in the $\Lambda$ continuum regions. 

\item[(2)] 
The two-pole structure for the $1^+_1$ and $1^+_2$ resonance states emerges 
near the ${^3{\rm He}}+d+\Lambda$ threshold, 
where interference effects between  ${^5{\rm Li}(3/2^+)}\otimes(0s_{1/2})_\Lambda$ 
and  ${^5{\rm Li}(1/2^+)}\otimes(0s_{1/2})_\Lambda$ modify the $\Lambda$ production spectrum, 
leading to a narrower peak at 13.8 MeV for $J^P=$ $1^+$. 

\item[(3)]
The magnitude of the differential cross sections with the Fermi-averaged 
$K^-n \to \pi^- \Lambda$ amplitude obtained through EOFA is approximately 
half that obtained using the SFA. 
Nevertheless, the absolute values of these cross sections agrees well 
with the experimental data, 
effectively resolving the discrepancy 
between the theories and BNL experimental results.

\end{itemize}

In conclusion, we have successfully explained the inclusive $\pi^-$ spectra 
for $\Lambda$ production 
in $^6_\Lambda$Li via the $^6$Li($K^-$,~$\pi^-$) reaction 
within the DWIA framework using EOFA, 
along with the Green's function method and $\Lambda$ folding-model potentials. 
Continuum effects, including nuclear spin structure considerations, 
enable accurate reproduction of the spectral data. 
This study provides a valuable framework for extracting essential information 
on the structure and production mechanisms of $\Lambda$ hypernuclei 
from experimental data. 
Further microscopic calculations of $\Lambda$-nucleus potentials, 
based on modern $\Lambda N$ interactions, are essential for gaining deeper 
insights into $YN$ and $YNN$ interactions that involve $YN$ short-range 
correlations~\cite{Kurihara82}. 
These aspects will be explored in future studies.  

\begin{acknowledgments}
The authors thank Dr.~A.~Dote, Professor Y.~Akaishi, and Professor S.~Shinmura 
for their valuable comments. This work was supported by Grants-in-Aid for 
Scientific Research (KAKENHI) from the Japan Society for the Promotion of Science: 
Scientific Research (C) (Grant No.~JP20K03954).
\end{acknowledgments}

%This work was supported by JSPS KAKENHI Grant Numbers 24105008, 25400278. 

%-------------------------------------------------------------------

\clearpage

%%%%%%%%%%%%%%%%%%%%%%%%%%%%%%%%%%%%%%%%%%%%%%%%%%%%%%%%%%%%%%%%
\end{document}